# Immunometabolism at the Crossroads of Infection: Mechanistic and Systems-Level Perspectives from Host and Pathogen

Sunayana Malla[1#], Nabia Shahreen[1#], and Rajib Saha[1*]

[1]Chemical and Biomolecular Engineering, University of Nebraska-Lincoln, Lincoln, NE, United States of America.

*Corresponding author; email: rsaha2@unl.edu

**Equal contributing authors

# Abstract:

The emerging field of immunometabolism has underscored the central role of metabolic pathways in orchestrating immune cell function. Far from being passive background processes, metabolic activities actively regulate key immune responses. Fundamental pathways such as glycolysis, the tricarboxylic acid (TCA) cycle, and oxidative phosphorylation critically shape the behavior of immune cells, influencing macrophage polarization, T cell activation, and dendritic cell function. In this review, we synthesize recent advances in immunometabolism, with a focus on the metabolic mechanisms that govern the responses of both innate and adaptive immune cells to bacterial, viral, and fungal pathogens. Drawing on experimental, computational, and integrative methodologies, we highlight how metabolic reprogramming contributes to host defense in response to infection. These findings reveal new opportunities for therapeutic intervention, suggesting that modulation of metabolic pathways could enhance immune function and improve pathogen clearance.

# Introduction:

Advancements in the study of immune cells have deepened our understanding of the complex processes that govern immune function[1]. While signaling cascades have long been considered the primary drivers of immune cell behavior and phenotypes, ongoing research has revealed that the story is far more intricate[2]. As we delve deeper into the mechanisms underlying immune responses, it has become clear that immune cells are highly heterogeneous, with each type exhibiting unique characteristics and functional roles[3–5]. For instance, macrophages once thought to be a uniform population are now recognized as highly plastic, encompassing multiple subtypes with distinct functions[6]. Similarly, neutrophils have been found to exhibit diverse behaviors depending on the stimuli and the surrounding microenvironment. These discoveries have also shed light on the dynamic interplay between immune cells and their interactions with pathogens [5]. Among the most significant developments in this area is the emergence of immunometabolism which explores the relationship between metabolic processes and immune function. This field has demonstrated that metabolic changes are not merely downstream effects of signaling pathways; rather, they can be key determinants of immune cell fate and function. Immunometabolism has thus become an essential framework for understanding immune responses and offers promising avenues for therapeutic intervention[7].

Immunity operates through several layers, most notably the innate and adaptive immune systems. Innate immunity involves cells such as macrophages, neutrophils, dendritic cells, mast cells, and other granulocytes. These cells serve as the body's first line of defense, responding rapidly to pathogens by attempting to neutralize them and initiating signals that activate the adaptive immune response[8,9]. The adaptive immune system, composed primarily of T cells and B cells, generates specific and long-lasting responses to pathogens, retaining immunological memory for future encounters[8,9]. Emerging evidence underscores the pivotal role of metabolic pathways in shaping immune responses, extending well beyond their traditional role in supplying energy. Metabolic activity not only fuels immune cell function but also actively regulates their activation, differentiation, and interaction with other cells and the surrounding microenvironment[7].

Foundational studies by Mathis & Shoelson. (2011)[10], Tannahill et al. (2013)[11], and Kidani et al. (2014)[12] highlight the importance of metabolism in modulating immune activity.

While early investigations primarily employed experimental approaches, they were often limited in scope, typically focusing on a narrow set of pathways or metabolites. In response, computational approaches have become essential for expanding the study of immune metabolism[13,14]. Genome-scale metabolic (GSM) modeling and machine learning (ML) have enabled a systems-level understanding of immune regulation, offering valuable insights into how metabolism governs immune responses[13,15]. However, it has become clear that the most comprehensive and impactful insights arise from integrated approaches, combining computational models with experimental validation [15]. This synergy allows researchers to explore complex metabolic networks in depth, generating hypotheses that are tested and refined through empirical methods, ultimately producing more holistic and reliable conclusions. Importantly, the metabolic behavior of immune cells is not static; it shifts in response to the type of pathogen encountered[13,15]. Whether confronting bacteria, viruses, or fungi, immune cells tailor their responses and underlying metabolic programs accordingly[16]. Thus, understanding these pathogen-specific metabolic shifts is crucial for deciphering immune responses and for identifying potential targets for therapeutic intervention.

In this review, we examine recent progress in the field of immunometabolism, with a particular focus on the contributions of experimental, computational, and integrative methodologies. Collectively, these approaches have highlighted key metabolic pathways, including glycolysis, oxidative phosphorylation (OXPHOS), the tricarboxylic acid (TCA) cycle, fatty acid oxidation (FAO), and amino acid metabolism as central regulators of immune cell function. While recent reviews Makowski et al. (2020)[1]; Purohit et al. (2022)[13]; and Basso et al., (2024)[17] have provided extensive analyses of metabolic processes in host immune cells, they have often overlooked pathogen-specific interactions and the distinct roles of various methodological approaches. Here, we address these gaps by exploring immune cell metabolic reprogramming during activation in both sterile and infectious contexts, with specific emphasis on bacterial, viral, and fungal pathogens. We delineate how each immune cell type exhibits distinct metabolic profiles and functional adaptations depending on the nature of the pathogen. The review further investigates how specific metabolic pathways underpin immune cell activation, differentiation, and effector functions. Moreover, we assess how recent insights have facilitated the discovery of novel metabolic targets for therapeutic modulation of immune responses. Particular attention is given to emerging roles of secondary metabolic pathways—such as arachidonic acid metabolism and leukotriene biosynthesis which are increasingly recognized as pivotal in immune regulation. As the field progresses, a comprehensive understanding of both primary and secondary metabolic circuits will be critical for elucidating immune dynamics and informing the development of precision immunotherapies.

# 1. Immunometabolism Advances in Host Immune cells

## 1.1. Innate Immune Cells

Advances in immunometabolism have highlighted how metabolic pathways such as glycolysis, OXPHOS, FAO, and TCA cycle are intricately tied to the function and plasticity of innate immune cells. Early experimental studies paved the way to understand and explore the true contribution of metabolic pathways in immune responses. Particularly, the role of central carbon metabolism has been extensively studied across multiple cell types and has been established as an important regulator during immune stimulation[18]. In macrophages, pro-inflammatory (M1) activation is accompanied by a metabolic shift toward glycolysis and a disrupted TCA cycle, enabling rapid energy production and support for inflammatory effector functions[19–21]. In contrast, alternatively activated M2 macrophages favor OXPHOS and FAO, sustaining their roles in tissue repair and immune regulation[22]. Dendritic cells (DCs) similarly upregulate glycolysis upon activation, a process essential for antigen presentation, cytokine production, and migration[23]. This shift is modulated by the mTOR pathway which balances glycolysis with mitochondrial respiration and is further influenced by the glucose availability[23,24]. DCs also utilize metabolic intermediates such as α-ketoglutarate to drive epigenetic changes that influence tolerogenic behavior[24]. Natural killer (NK) cells exhibit enhanced glycolytic and mitochondrial activity during viral infection, coupled with increased amino acid uptake, and iron availability[25]. Presence of iron supports NK effector functions, while iron deficiency impaired antiviral responses[26]. Similarly, Mast cells also rely on both glycolysis and OXPHOS to sustain degranulation and immune activation, further underscoring the role of dual metabolic engagement in effector responses[27,28]. However, in eosinophils, basal mitochondrial respiration and spare respiratory capacity are markedly high, reflecting a metabolism primed for sustained activity even in resting states[29]. Neutrophils on the other hand are marked by dominant glycolysis, and reduced number of mitochondria, a decrease in FAO, with an increase in the uptake of glucose. The shift in metabolic capacities of neutrophils is very distinct to different phases and activation states[30,31].

Beyond these core pathways, other metabolic programs add further layers of regulation. FAO is crucial in M2 macrophages and tolerogenic DCs, enabling longer-term immune functions, while glycogen metabolism in DCs provides a rapid energy reservoir during immune activation[25,32]. Lipid metabolism has also been implicated in cytokine production and DC migration in inflammatory contexts. Meanwhile, several amino acid metabolisms, particularly glutamine and arginine modulate macrophage polarization through nitric oxide synthesis and other downstream effects[33]. The tryptophan–kynurenine axis represents another regulatory mechanism, shaping immune responses via metabolite signaling, particularly in macrophages[31]. Accumulating evidence also highlights how metabolic intermediates such as itaconate, succinate, citrate, and α-ketoglutarate act not only as energy substrates but also as immune-signaling molecules that influence cell fate and function. Together, these insights reveal that innate immune responses are not simply influenced by metabolism, they are metabolically programmed. These experimental studies establish the interplay between glycolysis, mitochondrial function, and substrate availability that defines immune cell behavior and plasticity, making immunometabolism a promising target for modulating inflammation and host defense.

Additionally, computational tools have increasingly clarified how core and auxiliary metabolic pathways shape innate immune responses. Techniques such as genome-scale metabolic models (GSMs), flux balance analysis (FBA), and integrative multi-omics have enabled the simulation of glycolysis, OXPHOS, FAO, and TCA cycle activity under varying immune stimuli and activation states[34]. In addition to corroborating the experimentally reported phenomenon, computational analysis has enabled the in-depth analysis of metabolic networks beyond central carbon metabolism. For instance, models of alveolar macrophages have revealed subtype-specific enrichment in pyruvate metabolism, arachidonic acid pathways, and chondroitin/heparan sulfate biosynthesis and degradation—pathways that complement central carbon metabolism modules such as glycolysis, the pentose phosphate pathway (PPP), and the TCA cycle in defining macrophage polarization[35–38]. These computational frameworks extend beyond macrophages, offering insights into metabolic remodeling in dendritic cells (DCs), where glycolytic flux is predicted to rise during viral activation, as demonstrated in integrative analyses of gene expression and immune signaling networks following H1N1 and Newcastle Disease Virus exposure[39]. Complementary approaches incorporating transcriptomics, multiplex ELISA, qPCR, and flow cytometry have further mapped immune activation to metabolic shifts, including changes in TCA cycle flux and OXPHOS regulation in DCs[40]. Single-cell transcriptomics and network modeling have added further resolution, identifying tissue- and context-specific metabolic states that modulate antigen presentation and cytokine production in DC subsets[24]. These frameworks can predict key dependencies, including enhanced glycolysis in natural killer (NK) cells, under both physiological and pathological conditions[31].

Overall, computational modeling when paired with experimental validation has expanded our understanding of how specific metabolic pathways drive immune responses across disease contexts. Analyses of glycolysis, FAO, OXPHOS, and amino acid metabolism have proven central to modeling macrophage activation dynamics. For example, Russell et al. (2019) employed dual RNA sequencing across in vivo and in vitro infection settings, demonstrating that pathogen control is closely tied to macrophage metabolic polarization, with glycolysis dominating in pro-inflammatory states and FAO characterizing anti-inflammatory profiles[41]. These findings align with mechanistic models such as that developed by Zhao et al. (2021), which mapped metabolic control points in peripheral arterial disease, confirming experimentally that glycolysis supports M1 polarization, while FAO and OXPHOS facilitate M2 differentiation[42]. Broader systems-level reviews, such as Purohit et al. (2022), have emphasized how integrative approaches—including metabolomics, single-cell RNA sequencing, CRISPR screens, and metabolic flux analysis—continue to identify core dependencies in glycolysis, OXPHOS, and amino acid metabolism that shape immune cell fate and function[43]. Similar strategies have elucidated dendritic cell (DC) metabolic programs, with integrated modeling approaches converging on the critical role of glycolysis during activation and the modulatory contributions of FAO, lipid pathways, and TCA cycle intermediates to DC immune regulation. For example, Geeraerts et al. (2021) combined single-cell transcriptomics with pathway-level analysis to show how environmental cues reshape DC subset metabolism, thereby influencing their immunostimulatory versus tolerogenic outputs[24]. These findings collectively underscore the predictive power of computational models in capturing the metabolic plasticity underlying immune activation and regulation.

## 1.2. Adaptive Immune cells

Adaptive immune cells are specialized white blood cells that target specific pathogens and foreign substances, providing long-lasting immunity after exposure. Understanding the role of metabolism in the activation/maintenance of adaptive immune cells is very important and has been well studied via experimental approaches. T cell metabolism plays a pivotal role in regulating activation, differentiation, and immune function. Glycolysis serves as the primary energy source for effector T cells, enabling rapid responses during immune activation, while oxidative phosphorylation (OXPHOS) and fatty acid oxidation (FAO) support memory T cells' long-term survival and their ability to mount quick recall responses[44]. Levine et al. (2021) highlighted that early-activated CD8+ T cells display a distinct metabolic signature, marked by increased glycolytic enzyme levels (such as GAPDH and Glut1) alongside elevated mitochondrial proteins like ATP5A, which underscores their reliance on glycolysis and mitochondrial function[45]. The mitochondrial folate pathway also plays a crucial role in activated T cells, with Ron-Harel et al. further elucidating this mechanism[46]. Additionally, stable isotope labeling (SIL) studies have shown that effector CD8+ T cells *in vivo* exhibit low lactate production, relying extensively on glucose oxidation through the TCA cycle. Interestingly, pyruvate enters the TCA cycle via pyruvate dehydrogenase *in vivo*, while *in vitro*, pyruvate carboxylase plays a more significant role. Moreover, during the effector phase, serine biosynthesis becomes increasingly prominent, further highlighting the metabolic shifts that occur in activated T cells[47].

Though the metabolic regulation of B cells is less understood compared to other lymphocyte populations, recent research has shed light on the specific metabolic pathways that govern B cell functions. Capasso et al. (2015) demonstrated the critical role of glycolysis in B cell activation, yet there remain significant gaps in the broader metabolic framework that regulates B cell responses[48]. More recent work has found that activated B cells upregulate glycolysis to support their proliferation and antibody production, while regulatory B cells (Bregs) predominantly rely on OXPHOS for their functions. This metabolic divergence aligns with the distinct roles of effector and regulatory B cells in immune responses and disease modulation[49]. Rosser et al. (2021) further emphasized that OXPHOS is a defining characteristic of Breg activity. Computational models suggest that shifting metabolic pathways, such as from glycolysis to OXPHOS, could reprogram B cell functions, offering potential therapeutic strategies for diseases like autoimmunity, cancer, and infection[50].

The use of computational tools has significantly advanced our understanding of metabolic reprogramming in T cells, particularly in the context of regulatory T cells (Tregs), which play a crucial role in maintaining immune tolerance. Puniya et al. (2018) applied computational modeling to explore the complex interactions between cytokines, intracellular signaling molecules, and transcription factors that govern Treg behavior, revealing distinct metabolic signatures specific to this subset[51]. These integrative approaches offer a systems-level view of the regulatory networks influencing T cell metabolism and their immune functions. Ma et al. (2024) highlighted the critical role of serine metabolism in modulating mTORC signaling pathways, which are central to determining T cell fate. This study adds to the growing body of evidence suggesting that manipulating metabolic pathways can precisely direct immune responses[52]. Additionally, Sen et al. (2021) combined genome-scale metabolic modeling with transcriptomic and lipidomic data to demonstrate how human CD4+ T cells undergo distinct metabolic adaptations during activation and differentiation. Their findings underscore the pivotal roles of ceramide and glycosphingolipid biosynthesis in supporting Th17 lineage commitment and function [49]. Collectively, these studies illustrate the potential of integrating computational and experimental approaches to dissect and potentially modulate T cell metabolism in various immunological contexts.

## 2. Immunometabolism Advances in Pathogen-Specific Interactions

### 2.1. Viral Infections

Viruses universally reprogram host cell metabolism to support replication. Experimental studies have demonstrated that many viruses, including SARS-CoV-2 and HIV-1, induce significant metabolic shifts towards aerobic glycolysis, analogous to the Warburg effect[53]. SARS-CoV-2 infection increases glucose uptake, lactate production, and lipid metabolism in monocytes and macrophages via HIF-1α stabilization triggered by reactive oxygen species (ROS)[54]. This metabolic reprogramming provides essential biosynthetic intermediates for viral replication while suppressing antiviral interferon responses. Similarly, HIV-1 infection elevates glycolytic flux and mitochondrial biogenesis in CD4+ T cells and macrophages, driven primarily by HIF-1α, to meet energetic and biosynthetic demands for virion production[55,56].

Computational analyses employing GSM modeling have further clarified these interactions. Modeling of SARS-CoV-2-infected lung epithelial cells revealed significant disruptions in lipid metabolism, accurately predicting viral replication dynamics and key host metabolic vulnerabilities[57]. Similar computational approaches for influenza and HIV infections have identified essential host metabolic reactions exploited by viruses,

emphasizing glycolysis and nucleotide biosynthesis pathways as potential targets for antiviral therapy[58]

Integrative studies combining computational modeling and multi-omics validation have robustly confirmed host metabolic vulnerabilities in viral infections. Multi-omics analysis of COVID-19 patient samples confirmed glycolytic shifts and hyperactivation of the mTORC1 signaling pathway, consistent with computational predictions[59]. Pharmacological inhibition of glycolysis with compounds such as 2-deoxy-D-glucose (2-DG) has been experimentally validated to restrict viral replication across diverse virus families, confirming integrative computational predictions of metabolic chokepoints[60].

## 2.2. Bacterial Infections

While viruses exploit host metabolism to facilitate replication, bacterial pathogens employ distinct yet equally sophisticated strategies to reprogram host metabolic pathways, often engaging in dynamic metabolic "arms races" with immune cells. Experimental investigations have demonstrated that intracellular bacteria like *Mycobacterium tuberculosis* (Mtb) and *Salmonella enterica* serovar Typhimurium substantially alter macrophage metabolism. Classically activated macrophages infected with Mtb exhibit enhanced glycolysis and reduced oxidative phosphorylation, essential for IL-1β and nitric oxide-mediated bacterial clearance. However, Mtb actively counteracts this protective glycolytic shift by upregulating microRNA-21, suppressing critical glycolytic enzymes such as PFKFB3, and limiting host immune responses[61]. Furthermore, Mtb infection modulates macrophage iron metabolism through nitric oxide regulation of ferroportin, impacting pathogen survival[62]. Salmonella similarly disrupts host metabolism via effector proteins like SopE2, which inhibits serine biosynthesis, causing the accumulation of the glycolytic intermediate 3-phosphoglycerate (3PG) utilized by bacteria for replication. Additionally, Salmonella persistence in macrophages depends on host fatty acid oxidation driven by PPARδ signaling[63,64]. This metabolic ingenuity extends to sexually transmitted and antibiotic-resistant pathogens. *Treponema pallidum* scavenges host cholesterol to stabilize its membrane and evade immune detection, while *Neisseria gonorrhoeae* upregulates anaerobic respiration in oxygen-limited mucosal environments, enhancing persistence despite antibiotic pressure (scavenges host cholesterol)[65,66]. Similarly, methicillin-resistant *Staphylococcus aureus* hijacks host fatty acid oxidation pathways to counteract metabolic inhibitors, and *Pseudomonas aeruginosa* exploits nucleotide biosynthesis and biofilm-associated polysaccharides (e.g., alginate, Psl) to sustain chronic infections[67,68].

Complementing these experimental insights, computational modeling, particularly GSM models, has provided mechanistic insights into bacterial metabolic reprogramming strategies[69–71]. The pioneering metabolic model by Bordbar et al. of Mtb-infected macrophages identified critical pathogen-induced shifts in host metabolic pathways, providing an experimental roadmap for targeted metabolic intervention[35]. Integrated dual-host-pathogen transcriptomic models for Salmonella and Mtb infections have further pinpointed host metabolic vulnerabilities and pathogen-dependent metabolic chokepoints that may be exploited therapeutically[72].

Moreover, integrative approaches combining computational predictions with experimental validations have substantially deepened the understanding of bacterial-host metabolic interactions. Proteomic analysis integrated with flux modeling in Mtb-infected macrophages validated predicted glycolytic and amino acid metabolism perturbations, identifying specific metabolic nodes for therapeutic targeting[61]. Similarly, multi-omics and genetic screens in Salmonella infections identified itaconate accumulation as a critical antimicrobial host metabolite, strengthening integrative computational predictions regarding pathogen vulnerabilities[64,73].

## 2.3. Fungal and Parasitic Infections

Beyond bacterial and viral infections, eukaryotic pathogens such as fungi and parasites have evolved unique mechanisms to subvert host metabolism, often directly competing for nutrients or reprogramming immune cell metabolic states to establish chronic infections. Experimental evidence shows fungal pathogens like *Candida albicans* induce distinct immunometabolic shifts within host immune cells. Candida cell-wall β-glucans trigger long-term trained immunity in monocytes by shifting cellular metabolism from oxidative phosphorylation to glycolysis via mTOR and HIF-1α-dependent pathways, enhancing innate immune responses upon pathogen rechallenge [74,75]. Moreover, Candida adapts metabolically by utilizing alternative carbon sources during host-imposed nutrient restrictions, underscoring pathogen metabolic plasticity[76].

Parasitic pathogens similarly manipulate host metabolism. Experimental studies on *Toxoplasma gondii* demonstrate host-driven nutrient competition, as interferon-γ-induced indoleamine 2,3-dioxygenase (IDO) depletes tryptophan, restricting parasite growth[77]. Toxoplasma also manipulates host mitochondria to siphon fatty acids, further elucidating the role of nutrient competition in parasitic infections[78]. Leishmania parasites exploit host macrophage arginine metabolism, diverting it to polyamine synthesis via arginase-1 activity and diminishing nitric oxide production required for parasite elimination[79,80]. Schistosoma mansoni similarly modulates host immune metabolism by driving Th2 responses and M2 macrophage polarization, characterized by increased arginase-1 expression, promoting parasite persistence within granulomas[81].

Here too, computational modeling has identified critical metabolic vulnerabilities in fungal and parasitic pathogens. Genome-scale metabolic reconstructions like CandidaNet for Candida albicans pinpoint essential amino acid biosynthesis pathways critical for pathogen survival, presenting targeted therapeutic opportunities[82]. Similarly, ToxoNet1, a comprehensive metabolic model of Toxoplasma gondii, identified synthetic lethal gene interactions essential for parasite survival, guiding experimental validation[83].

Finally, integrative approaches merging computational predictions with multi-omics validation have significantly advanced understanding of host-pathogen interactions. Multi-omics analysis of β-glucan-trained monocytes experimentally validated long-term epigenetic and metabolic rewiring predicted by computational models, reinforcing glycolysis-driven trained immunity as a therapeutic target against fungal infections[84]. Likewise, integrative studies involving genomic, transcriptomic, and metabolic modeling of Leishmania and Toxoplasma have confirmed critical host and parasite metabolic chokepoints, providing mechanistic bases for therapeutic intervention strategies[80,83].

## 3. Therapeutic Directions

### 3.1. Host-Directed Metabolic Therapies

Advances in immunometabolism have spurred host-directed therapies that reprogram the host's metabolism to combat infection. One strategy is to inhibit glycolysis, a pathway often hijacked by both immune cells and pathogens during infection. 2-Deoxy-D-glucose (2-DG), a glucose analog that blocks hexokinase, has been shown to dampen hyperinflammatory responses in viral and bacterial infections while preserving essential immune functions[85]. By curbing the "Warburg-like" glycolytic shift in activated immune cells, 2-DG can reduce damaging cytokine storms without completely impairing host defense[85]. Metformin, an AMPK activator widely used for type 2 diabetes, is another prominent host-directed agent. Metformin skews immune metabolism away from glycolysis toward oxidative phosphorylation, which enhances the microbicidal capacity of macrophages[86]. In Mtb infection, metformin triggers mitochondrial reactive oxygen species production and augments autophagy, thereby restricting intracellular Mtb growth[86]. Notably, diabetic patients on metformin have shown improved tuberculosis outcomes, and metformin is under investigation as an adjunct to standard anti-TB therapy. Arginine supplementation has likewise been explored to bolster host immunity. L-arginine is a semi-essential amino acid needed for nitric oxide (NO) synthesis, a key antimicrobial effector molecule. Clinical trials in pulmonary tuberculosis have yielded mixed results: one study reported faster symptom resolution and weight gain with arginine adjunct therapy, presumably via enhanced NO production[87], whereas another trial found no significant improvement in sputum clearance or clinical outcomes[88]. These discrepancies underscore the importance of optimizing dosage and patient selection for metabolic adjuncts.

Statins, classically used to lower cholesterol, have emerged as promising immunometabolic adjuvants that modulate host lipid pathways and inflammation. By inhibiting HMG-CoA reductase, statins not only reduce cholesterol (a nutrient some pathogens scavenge) but also exert direct immunomodulatory and antimicrobial effects. For example, simvastatin treatment enhanced macrophage bactericidal activity against Mtb by promoting autophagy and phagosome maturation[89]. In mice infected with Mtb, adjunctive simvastatin shortened the duration of antibiotic therapy needed to sterilize lungs, suggesting that host lipid manipulation can accelerate pathogen clearance[90]. Early-phase clinical studies are now examining statins as adjuncts in TB and other infections to determine if these preclinical benefits translate to humans. Importantly, the safety profile of host-directed therapies must be carefully considered – broad metabolic inhibitors like 2-DG or AMPK activators can have off-target effects on tissues with high metabolic demands. Nonetheless, these approaches illustrate the therapeutic potential of tipping host metabolic balances to favor antimicrobial immunity.

With all the combined efforts, several therapeutic targets have emerged due to advancements in immunometabolism. Manipulating metabolic pathways, such as glycolysis, lipid metabolism, or glutaminolysis, can alter the phenotype and function of immune cells, potentially shifting them from pro-inflammatory to anti-inflammatory states[91]. Some therapeutic agents, like dimethyl fumarate (DMF) and metformin, have been shown to affect immune cell metabolism and have anti-inflammatory effects[92]. Specific T cell subsets, like Th17 cells, utilize glycolysis for energy, while others, like regulatory T cells (Tregs), rely on lipid metabolism. Targeting these pathways can influence T cell function and potentially modulate immune responses[93].

### 3.2. Targeting Pathogen Metabolic Vulnerabilities

Contrary to the host-directed therapies, exploiting specific metabolic weaknesses of pathogens has also become an attractive strategy to develop targeted therapies. A quintessential example is the frontline tuberculosis drug isoniazid, which inhibits the enoyl-ACP reductase (InhA) in the mycobacterial fatty acid synthesis pathway. This blockade prevents Mtb from synthesizing mycolic acids, essential components of its cell wall, thereby killing the bacterium. Isoniazid's success showcases how disabling a pathogen's unique metabolic enzyme can have potent therapeutic effects. Researchers are extending this concept by identifying other pathogen-specific metabolic bottlenecks. One host-derived metabolite, itaconate, has drawn considerable interest for its antimicrobial properties. Activated macrophages naturally produce itaconate via IRG1 (ACOD1) as a defensive strategy. Itaconate can inhibit pathogens by targeting key enzymes, for instance, itaconate is a competitive inhibitor of isocitrate lyase in Mtb, crippling the glyoxylate shunt that Mtb relies on for persistence during nutrient stress[73]. Building on this innate mechanism, synthetic itaconate analogs (such as 4-octyl-itaconate and others) are being developed to therapeutically exploit this pathway. These cell-permeable analogs aim to deliver high concentrations of itaconate or mimic into infected tissues, thereby directly suppressing bacterial growth and simultaneously inducing anti-inflammatory Nrf2 signaling in host cells. Preclinical studies of itaconate derivatives in models of sepsis and viral infection have already demonstrated reduced inflammatory damage and improved pathogen control, highlighting their dual action on host and microbe.

Auxotrophies of pathogens, which are nutritional dependencies resulting from their inability to synthesize certain metabolites, offer another set of targets for intervention. Many intracellular pathogens must scavenge amino acids from the host, and the host immune system can exploit this by depriving them of these nutrients. Toxoplasma gondii, for example, is auxotrophic for tryptophan: it cannot synthesize this essential amino acid and must acquire it from host cells. The host capitalizes on this vulnerability through the interferon-gamma–induced enzyme indoleamine 2,3-dioxygenase (IDO), which degrades tryptophan in infected cells. This mechanism starves T. gondii and inhibits its replication[94]. Analogous auxotrophies exist for arginine in some pathogens, and therapies enhancing or mimicking these nutrient-withholding tactics are being explored. For instance, recombinant IDO or small-molecule IDO inducers could, in theory, be used to reinforce tryptophan starvation as an anti-Toxoplasma strategy, though care must be taken given IDO's broader role in immune regulation. Similarly, enzymes that degrade arginine or other critical nutrients in the infection microenvironment might suppress microbial growth, although they risk impairing host cells that require the same nutrients.

Targeting pathogen metabolism also includes repurposing existing drugs that incidentally affect microbial nutrient acquisition. As noted, statins not only act on the host but also limit the availability of cholesterol and isoprenoids that certain bacteria and parasites need for their membranes and virulence factor production. Some extracellular bacteria and protozoa are unable to synthesize sterols or isoprenoid precursors and must obtain them from the host; statin-mediated depletion of these metabolites can thus directly attenuate pathogen growth while modulating host immunity. Additionally, antifolate and antimetabolite drugs (like trimethoprim or sulfa drugs) exemplify the targeting of microbial vitamin and nucleotide synthesis, although these classical examples go beyond immunometabolism per se. The unifying theme is that detailed mapping of pathogen metabolic pathways can reveal choke points that are absent or dispensable in the host, allowing for precision strikes with minimal toxicity. Modern metabolomic and genetic approaches are rapidly expanding the list of such pathogen-specific vulnerabilities. Together, host-directed and pathogen-targeted strategies demonstrate the translational promise of immunometabolic interventions in infectious diseases. Several candidates have advanced to clinical trials, but further optimization of dosing, combination strategies, and mechanistic targeting is essential to maximize therapeutic efficacy and minimize risk.

## 4. Future Direction & Our Perspective

While current research in immunometabolism has greatly expanded our understanding of central carbon metabolism in host defense, substantial gaps remain in our grasp of secondary metabolic pathways and their context-dependent roles. Fatty acids, amino acid derivatives, and lipid mediators, such as prostaglandins and leukotrienes, profoundly influence immune cell differentiation, cytokine production, and inflammation resolution[1,92]. Yet, these secondary metabolites are often understudied in both experimental and computational frameworks, limiting our ability to capture the full regulatory landscape of immune metabolism. Emerging multi-dimensional tools are now beginning to address this complexity. Spatially resolved multi-omics platforms integrate transcriptomic, proteomic, and metabolomic data within intact tissues, revealing metabolically distinct immune microenvironments that shape infection outcomes. For example, imaging mass spectrometry coupled with spatial transcriptomics has begun to map immunometabolic heterogeneity in tumor and inflamed tissues, uncovering niches where distinct immune-metabolic programs dominate[95]. Such approaches, if applied to infectious disease models and organoid systems, could reveal when and where metabolic rewiring is protective versus pathogenic.

Alongside experimental advances, computational modeling must evolve to capture the bidirectional nature of host–pathogen metabolic interactions. Current genome-scale models often focus on host or pathogen metabolism in isolation. New community-level metabolic reconstructions, incorporating both immune cells and microbes, can simulate nutrient competition and predict synthetic-lethal vulnerabilities. For example, dual-organism models parameterized by transcriptomic and metabolomic data have successfully identified host–microbe nutrient bottlenecks, which could be exploited for precision antimicrobial therapies[96]. Moreover, metabolite-based biomarkers are emerging as powerful tools for patient stratification. Circulating levels of lactate, itaconate, or kynurenine correlate with disease severity in infections such as tuberculosis and COVID-19 and may help predict therapeutic response to metabolic modulators like metformin or 2-deoxy-D-glucose[97,98]. Incorporating these biomarkers into clinical trial design could reduce heterogeneity and enhance treatment efficacy.

Finally, the intersection of dietary inputs, microbiome composition, and immune metabolism offers untapped therapeutic potential. Short-chain fatty acids produced by commensals promote regulatory T cell differentiation and modulate innate immunity via epigenetic and signaling pathways[99]. Diet–microbiome–immune crosstalk is therefore not peripheral but central to immunometabolic regulation and may be harnessed to augment host defense or reduce inflammatory pathology. In summary, advancing immunometabolism from descriptive studies to mechanistically informed therapies will require the integration of spatial omics, bidirectional modeling, precision biomarkers, and microbial ecology. These tools promise to reshape our understanding of immune function in infection and inflammation, offering a roadmap for metabolic interventions that are targeted, personalized, and durable.

## 5. Conclusion

Immunometabolism has redefined our understanding of how immune cells sense, respond to, and shape their environment during infection. No longer viewed as passive energy consumers, immune cells actively rewire their metabolic programs to control effector function, proliferation, and differentiation. This reprogramming is context-dependent: glycolysis supports rapid cytokine production in pro-inflammatory states, whereas oxidative phosphorylation and fatty acid oxidation sustain long-lived or regulatory functions. Pathogens, in turn, adapt their metabolism to evade immune attack and exploit host-derived resources. From viral-induced glycolysis to bacterial manipulation of host serine biosynthesis and parasite-driven modulation of amino acid pathways, pathogens engage in a dynamic metabolic interplay with the immune system. These interactions not only determine infection outcomes but also shape systemic immune responses and inflammation. The convergence of experimental, computational, and integrative approaches has illuminated critical metabolic nodes, such as itaconate synthesis, lactate accumulation, and arginine catabolism that govern host–pathogen dynamics. These insights are now informing therapeutic development. Host-directed strategies that modulate immune metabolism, such as metformin or AMPK activators, and pathogen-targeted interventions that exploit metabolic bottlenecks, such as isoniazid or synthetic itaconate analogs, exemplify this translational promise. Looking ahead, the field is poised to move beyond single-cell models and isolated pathways. By combining spatial multi-omics, dual-organism modeling, microbiome-informed interventions, and metabolite-based patient stratification, immunometabolism offers a transformative framework for infectious disease research. Its integration into clinical practice may not only improve infection control but also help modulate immune responses in chronic inflammation, autoimmunity, and cancer.

## Conflicts of Interest
The authors declare no competing interests.

## Funding
National Institute of Health (NIH) R35 MIRA grant (5R35GM143009), awarded to RS.

## Acknowledgments
The figures were generated using bio render. and the meta-analysis was conducted by obtaining publication data from google scholar.

## Author contributions

S.M. and N.S. worked on concept development for this work and worked on writing and editing. The review and editing were done by R.S., S.M., and N.S.

**Figure Legends:**

Figure 1: The figure highlights the contribution of metabolism to boost immune response from solely host perspective and from the impact of different pathogens. The metabolic pathways highlighted in green denotes enhanced activity and red shows inhibited activity in each cell and pathogen type. The graph showing the tends of publication since year 2000 further highlights the rapid progression of immunometabolism field. The data for the graph is obtained from google scholar for term "immunometabolism" till May 2025.

**Figure 2:** Schematic showing possible topics that can be addressed to advance immunometabolism such as enhancing the study of secondary metabolism interactions within the host cells and the interaction with pathogens, which will ultimately aid development of novel and effective immunotherapeutic strategies.

**Figures:**

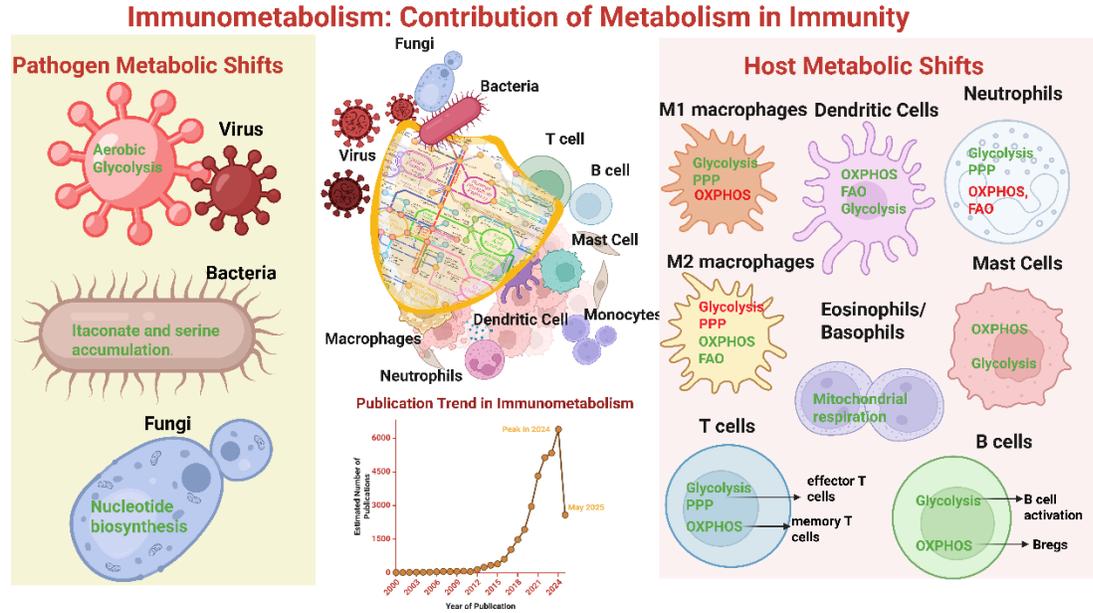

**Figure 2:**

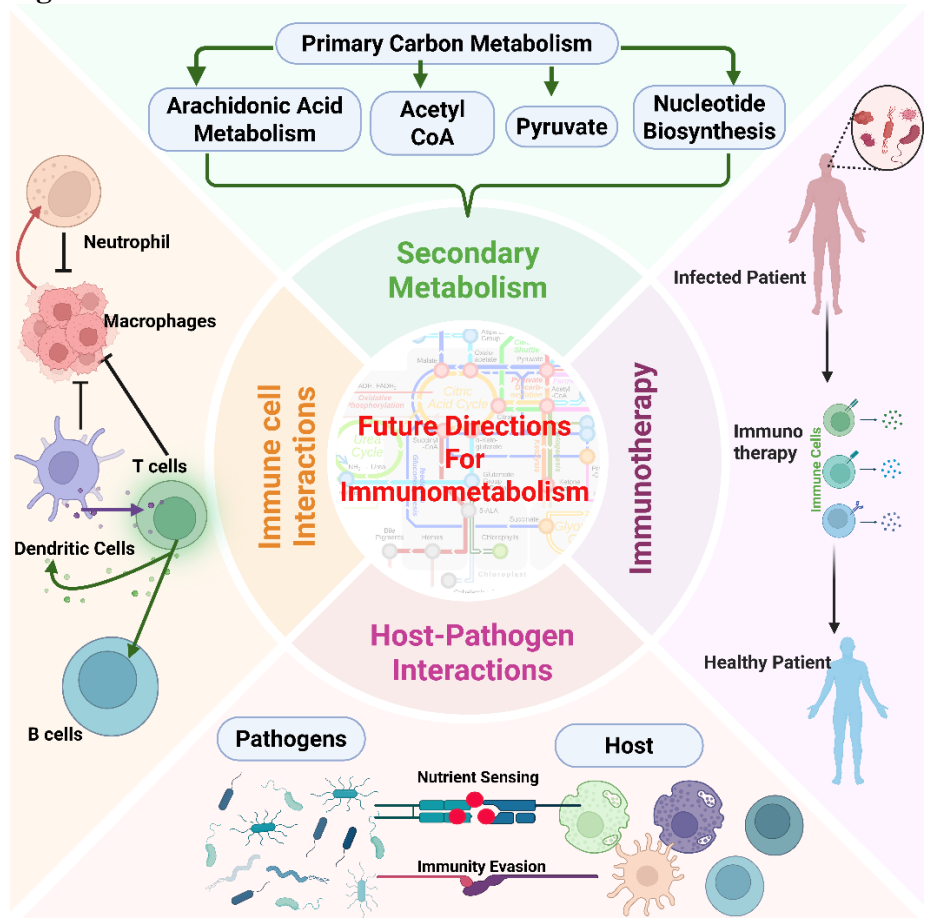

**Tables:**

Table 1: Table showing metabolic signatures of each immune cells and pathogens discussed throughout the paper.

| Immune Cell & Pathogens | Metabolic Signatures |
|---|---|
| M1 macrophages (Pro-inflammatory) | Increased glycolysis, disrupted TCA cycle[19–21], pyruvate metabolism, arachidonic acid metabolism, chondroitin/heparan sulphate biosynthesis, Pentose Phosphate Pathway [35–38] |
| M2 macrophages (Anti-inflammatory) | Increased OXPHOS, FAO[22], pyruvate metabolism, arachidonic acid metabolism, chondroitin/heparan sulphate degradation[35–38]. |
| Dendritic Cells | Increased glycolysis, increased utilization of α-ketoglutarate, increase glycogen metabolism, and Lipid metabolism, |

| | TCA and OXPHOS [23–25,32,39,40] |
|---|---|
| Neutrophils | Glycolysis dominant with increased PPP, decreased activity in OXPHOS, FAO, tryptophan–kynurenine[5,30,31] |
| Natural Killer cells | Increased glycolytic and mitochondrial activity, enhanced nutrient uptake, role of metals such as iron[75] |
| Mast Cells | Rely on both glycolysis and OXPHOS [11,27,28] |
| Other Granulocytes (Eosinophils, Basophils, etc.) | Increased basal mitochondrial respiration and spare respiratory capacity[29] |
| T cells | Increased glycolysis, elevated mitochondrial proteins, oxidative phosphorylation (OXPHOS) and fatty acid oxidation (FAO) support memory T cells and serine biosynthesis in activated T cells[45,55,77,94,99] |
| B cells | Upregulated glycolysis in B cell activation, Bregs rely in OXPHOS[48–50,50] |
| SARS-CoV-2 | Increased aerobic glycolysis (Warburg effect), increased glucose uptake lactate production, and lipid metabolism in monocytes and macrophages[36,37,54,57] |
| HIV-1 | Increased aerobic glycolysis & reactive oxygen species (ROS), elevated glycolytic flux and mitochondrial biogenesis in CD4+ T cells and macrophages[55] |
| Mycobacterium tuberculosis | Enhanced glycolysis/ reduced OXPHOS IN M1 MACROPAHGES, SUPPRESSES PFKFB3[61,62,88,89,97], effects iron metabolism through NO cycle |
| Salmonella enterica | Inhibited serine biosynthesis, accumulation of the glycolytic intermediate 3-phosphoglycerate (3PG), increased fatty acid oxidation, itaconate accumulation[64,96,96] |
| Treponema pallidum | Exploited host cholesterol metabolism.[70] |
| Neisseria gonorrhoeae | Upregulated anaerobic respiration [66,69] |
| Pseudomonas aeruginosa | Exploited nucleotide biosynthesis and biofilm-associated polysaccharides[68] |
| Staphylococcus aureus | Hijacked host fatty acid oxidation[67,71] |